\documentclass{article} 
\usepackage{GEM_workshop_2025,times,graphicx,subcaption}

\usepackage{amsmath,amsfonts,bm}

\def\eqref#1{equation~\ref{#1}}

\def\1{\bm{1}}

\DeclareMathAlphabet{\mathsfit}{\encodingdefault}{\sfdefault}{m}{sl}
\SetMathAlphabet{\mathsfit}{bold}{\encodingdefault}{\sfdefault}{bx}{n}

\usepackage{hyperref}
\usepackage{url}

\title{Non-Canonical Crosslinks Confound \\ Evolutionary Protein Structure Models}

\author{Romain Lacombe \\
Stanford University \\
\texttt{rlacombe@stanford.edu}
}

\iclrfinalcopy 

\begin{document}

\maketitle

\begin{abstract}
Evolution-based protein structure prediction models have achieved breakthrough success in recent years. However, they struggle to generalize beyond evolutionary priors and on sequences lacking rich homologous data. Here we present a novel, out-of-domain benchmark based on sactipeptides, a rare class of ribosomally synthesized and post-translationally modified peptides (RiPPs) characterized by sulfur to $\alpha$-carbon thioether bridges creating cross-links between cysteine residues and backbone. We evaluate recent models on predicting conformations compatible with these cross-links bridges for the 10 known sactipeptides with elucidated post-translational modifications. Crucially, the structures of 5 of them have not yet been experimentally resolved. This makes the task a challenging problem for evolution-based models, which we find exhibit limited performance (0.0\% to 19.2\% GDT-TS on sulfur to $\alpha$-carbon distance). Our results point at the need for physics-informed models to sustain progress in biomolecular structure prediction.

\end{abstract}

\section{Introduction}

Evolution-based protein structure prediction models have achieved breakthrough success in recent years, leading organizers of the Critical Assessment of Structure Prediction (CASP) \citep{kryshtafovych2021casp14}, a longstanding bi-annual blind prediction contest, to announce the protein folding problem as ``solved" for single-chain proteins \citep{StraitonSolved}. The authors of RoseTTA Fold \citep{baek2021} and AlphaFold \citep{jumper2021} were awarded the 2024 Nobel Prize in Chemistry for their work on protein design and protein structure prediction \citep{nobel_chemistry_2024}.

\begin{figure}[!b]
    \centering
    \includegraphics[width=0.6\linewidth]{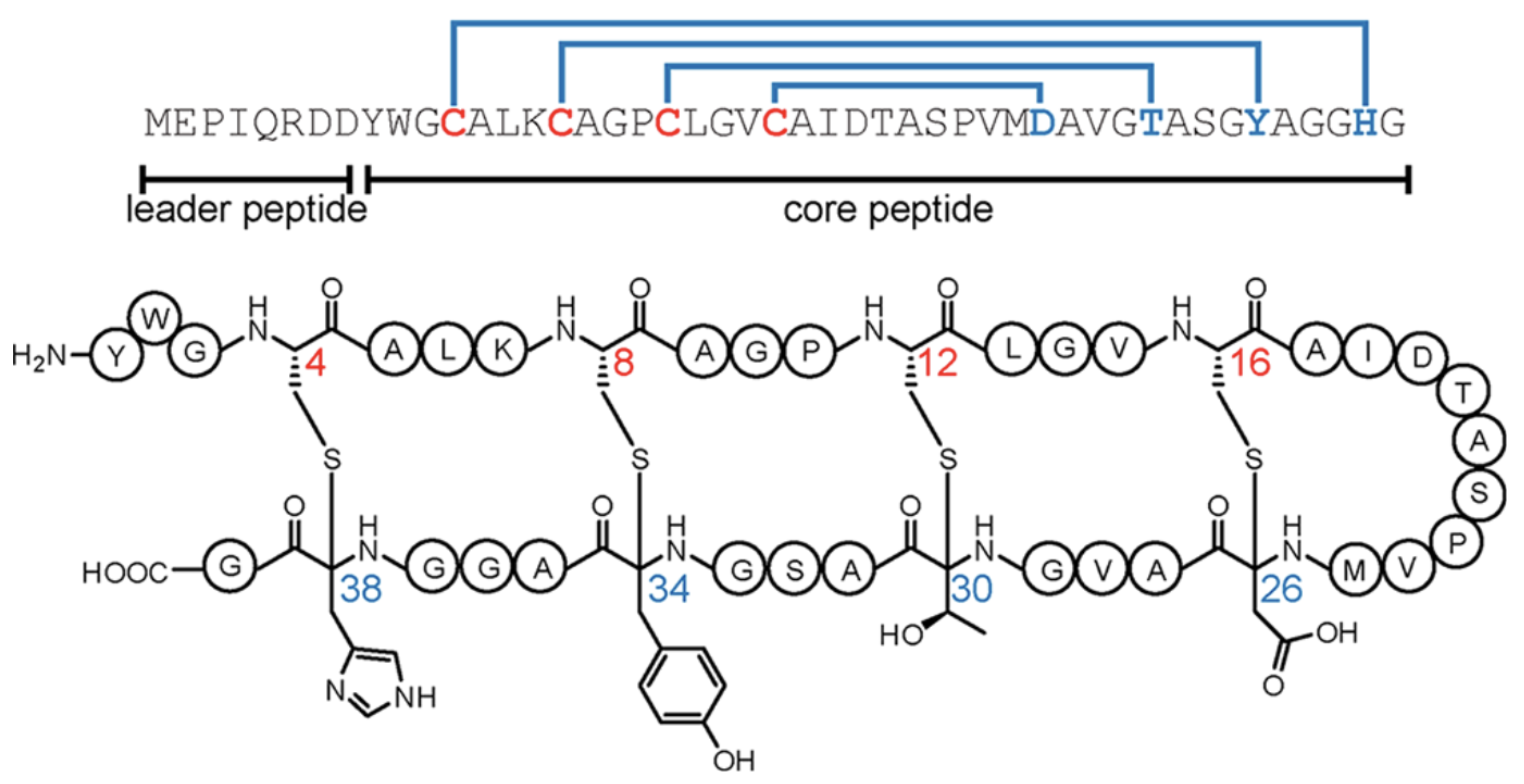}
    \caption{Example of sactipeptide (huazacin) and its post-translationally modified sulfur-to-alpha-carbon thioether bonds (blue nested hairpins). Figure from \citet{Huazacin}.}
    \label{fig:sactibonds}
\end{figure}

Despite these major achievements, evolution-based methods, which rely on multiple sequence alignments (MSAs) or evolutionary scale pre-training, generally struggle with out-of-domain proteins for which little or no homologous data is available. They are not trained to account for the dynamics of peptide chains, which, compared to physics-based approaches, limits their ability to generalize and their performances in areas such as allosteric control, \emph{de novo} peptide design, prediction of mutation effects, or complex interactions with other macro-molecules \citep{notin2023proteingym}.  

Evolution-based methods primarily train models on proteins with rich homology to known structures from the Protein Data Bank (PDB) \citep{ProteinDataBank}, and benchmarks such as CASP assess them on peptides for which experimental structures have recently been resolved. Proteins with complex post-translational modifications (PTMs), especially rare ones for which little data is available, remain a challenge for structure elucidation, and therefore for protein structure prediction. 

Sactipeptides are such a rare class of ribosomally synthesized and post-translationally modified peptides (RiPPs), characterized by their unique \textbf{sactionine bonds} or \textbf{`sactibonds'}, a thioether crosslink between the sulfur in a cysteine residue and a nearby $\alpha$-carbon atom in the backbone \citep{RecentAdvances2022}. These post-translation modifications are crucial for the biological function of sactipeptides, yet they remain poorly represented in training data. \textbf{Their rarity makes them an ideal benchmark for testing the robustness of protein structure prediction models beyond evolutionary priors.}

In this work, we introduce a novel benchmark to evaluate the out-of-domain performance of protein structure prediction models. We define a zero-shot task: predicting the 3D cross-linking structure of sactipeptides, specifically the distance between the sulfur and $\alpha$-carbon atoms forming sactibond bridges. We evaluate recent protein models on this task over \textbf{the only 10 known sactipeptides} with elucidated thioether cross-links, only 5 of which have an experimentally resolved structure, providing a challenging, rigorous assessment of their ability to generalize beyond their training set.

\section{Related Works}

\subsection{Protein Structure Prediction Models}
Deep learning has transformed protein structure prediction. AlphaFold2 achieved near-experimental accuracy by leveraging co-evolutionary signals through multiple sequence alignments (MSAs) \citep{jumper2021}. RoseTTAFold introduced a three-track network that integrates sequence, MSA, and pairwise features \citep{baek2021}. Other approaches such as ESMFold \citep{lin2023} and OmegaFold \citep{wu2022omegafold} eliminate the need for MSAs entirely by employing large-scale protein language models, achieving rapid inference with competitive accuracy. Open-source models provide efficient and accessible alternatives, such as such as ColabFold \citep{mirdita2022}, OpenFold \citep{bouatta2023openfold}, and Boltz-1 \citep{wohlwend2024boltz1}. 

The reliance of deep learning models on evolutionary signals still limits their applicability to proteins with poor sequence homology. Physics-based models such as Rosetta \citep{leaverfay2011rosetta3} and molecular dynamics (MD) simulations \citep{lindorff2011} remain essential for modeling noncanonical protein structures, intrinsically disordered regions, and cases where evolutionary priors are absent. Future progress will likely require integrating deep learning with molecular physics.

\subsection{Benchmarking Protein Prediction}
The CASP (Critical Assessment of Structure Prediction) experiment has been the gold standard for benchmarking models for three decades \citep{moult1995, kryshtafovych2021casp14}. The dominance of AlphaFold~2’s in CASP14 underscored the maturity of deep learning for single-domain protein folding. However, challenges remain in modeling multi-chain complexes, alternative conformations, protein-ligand interactions, and mutation impact prediction. To address these, new benchmarks such as ProteinBench \citep{ye2023proteinbench} and ProteinGym \citep{notin2023proteingym} evaluate models on broader tasks, including fitness prediction and protein design, to encourage generalization and continued progress in the field.

\subsection{Prediction Post-Translational Modifications}
Recent deep learning techniques have demonstrated impressive gains in PTM site prediction for common modifications such as phosphorylation, glycosylation, and methylation \citep{zhang2024deepo, wang2020musitedeep}. For RiPPs specifically, tools such as RiPPMiner \citep{Agrawal2017-RiPPMiner} and antiSMASH \citep{Blin2023-antiSMASH7} integrate genome mining with rule-based or machine learning pipelines to identify possible RiPP gene clusters and predict PTM topologies, bridging the gap between sequence data and experimentally validated structures. 



\section{Methods}

\subsection{Dataset: Sactipeptides Sequences and Cross-links}

Sactipeptides---short for sulfur-to-$\alpha$-carbon thioether-containing peptide---are a small but growing subclass of RiPPs natural products characterized by one or more intramolecular thioether linkages known as \textbf{sactionine bonds}. These unique cross-links are formed when a radical S-adenosylmethionine (rSAM) enzyme facilitates the covalent bonding of the sulfur atom of a cysteine residue to the $\alpha$-carbon of another amino acid in the peptide backbone. The result is a tightly \textbf{cross-linked polycyclic peptide} in which the \textbf{thioether bridges structure} from residue to backbone imparts rigidity and extreme stability against heat, pH, and proteases \citep{FluheSactipeptide}.

While the first sactipeptide---subtilosin A, an antibiotic produced by Bacillus subtilis 168---was discovered in 1985 \citep{Sactipeptide1985}, this class of RiPPs is still very rare, with the pace of discovery only ramping up in recent years thanks to advances in genome mining  \citep{CurrentAdvancements, RecentAdvances2022, GenomeMining}. A literature search reveals that to date, only 10 sactipeptides have a known sequence and fully elucidated cross-links structure. Of these, only 5 sactipeptides---ruminococcin C1 \citep{, RumC1}, subtilosin A \citep{SubtilosinA}, thurincin H \citep{ThurincinH}, thuricin CD $\alpha$, and thuricin CD $\beta$ \citep{ThuricinCD}---have an experimentally resolved 3D structure available in the PDB. To the best of our knowledge, the remaining 5 sactipeptides---huazacin \citep{Huazacin}, hyicin 4244 \citep{Duarte2018Hyicin}, sporulation killing factor A \citep{skfA}, streptosactin \citep{CurrentAdvancements}, and QmpA \citep{QmpA}---do not have a known structure. We lists these 10 sactipeptides and their post translational cross-links in table \ref{tab:sactipeptides}. 

Because half of these peptides are present in the PDB, and the other half have \textbf{identified cross-links} but \textbf{\textit{not yet} an experimentally resolved 3D structure}, they form an ideal held-out dataset for an out-of-domain evaluation of the robustness of protein structure prediction models.

\begin{table}[tbp]
    \centering
    \begin{tabular}{lccl}
       \textbf{Sactipeptide} & \textbf{Length} & \textbf{PDB ID} & \textbf{Cross-links} \\
       \hline \\
Ruminococcin C1 & 44 & 6T33 & C3→N16, C5→A12, C22→K42, C26→R34 \\
Subtilosin A  & 35 & 1PXQ & C4→F31, C7→T28, C13→F22 \\
Thurincin H & 31 & 2LBZ & C4→S28, C7→T25, C10→T22, C13→N19 \\
Thuricin CD $\alpha$ & 30 & 2L9X & C5→T28, C9→T25, C13→S21\\
Thuricin CD $\beta$ & 30 & 2LA0 & C5→Y28, C9→A25, C13→T21 \\
Huazacin    & 40 & --- &  C4→H38, C8→Y34, C12→T30, C16→D26 \\
Hyicin 4244 & 35 & --- & C4→F31, C7→T28, C13→F22 \\
Skf A  & 26 & --- & C4→M12 \\
Streptosactin  & 14 & --- & C4→S7, C10→G13 \\
QmpA & 13 & --- & C6→D4, C10→D8 \\
    \end{tabular}
    \caption{Summary of characterized sactipeptides, lengths of mature peptides, PDB entry if the structure is resolved, and post-translational cross-links structures.}
    \label{tab:sactipeptides}
\end{table}

\subsection{Cross-link Structure Prediction Task}

Our goal in this work is to probe protein structure prediction models to evaluate whether they can predict a conformation consistent with post-translational cross-links observed on sactipeptides. Crucially, \textbf{the length of sulfur-to-$\alpha$-carbon bonds is known}, measuring 1.8 Å in structures reported in the PDB. Therefore, knowing which cysteine residues form sactibonds with which target $\alpha$-carbon atoms gives us \textbf{critical geometric information} on the 3D structure of the peptide. 

We measure the predicted distances between S and C$_\alpha$ atoms in known sactibonds, and how closely they match this experimentally known bond length, to assess how well structures predicted by the different protein models respect the geometric bond length constraints the true structure must verify.

\textbf{Metric: GDT-TS}. The global distance test total score quantifies how many of the predicted geometries of known crosslink bonds match the experimentally observed sactibond length by less than 1 \AA, 2 \AA, 4 \AA, or 8 \AA 
 error, with GDT-TS of 100\% denoting perfect crosslink geometry prediction:

\begin{equation}
    \text{GDT-TS} = \frac{1}{4} \sum_{D \in \{1,2,4,8\}} \% \{\text{sactibond} \ | \ d(\textup{S}, \textup{C}^{\textrm{target}}_\alpha) \leq D + 1.8 \textup{~\AA} \}
\end{equation}

\textbf{Metric: RMSD}. This metric quantifies how far the model’s predictions are from aligning with experimentally observed sactibond geometries. The root mean square distance (RMSD) is computed between the predicted and experimentally validated sulfur-to-$\alpha$-carbon bond distances as follows, with an RMSD of 0 \AA\ denoting a perfect crosslink geometry prediction:

\begin{equation}
    \textrm{RMSD} = \sqrt{ \sum_{\text{S, C} \in \textrm{sactibonds}} || d(\textup{S}, \textup{C}^{\textrm{target}}_\alpha) - 1.8 \textup{~\AA}   ||^2 }  
\end{equation}

\subsection{Experiments}

\begin{figure}[!hb]
    \centering
    \begin{subfigure}{0.49\textwidth}
        \centering
        \includegraphics[width=\textwidth]{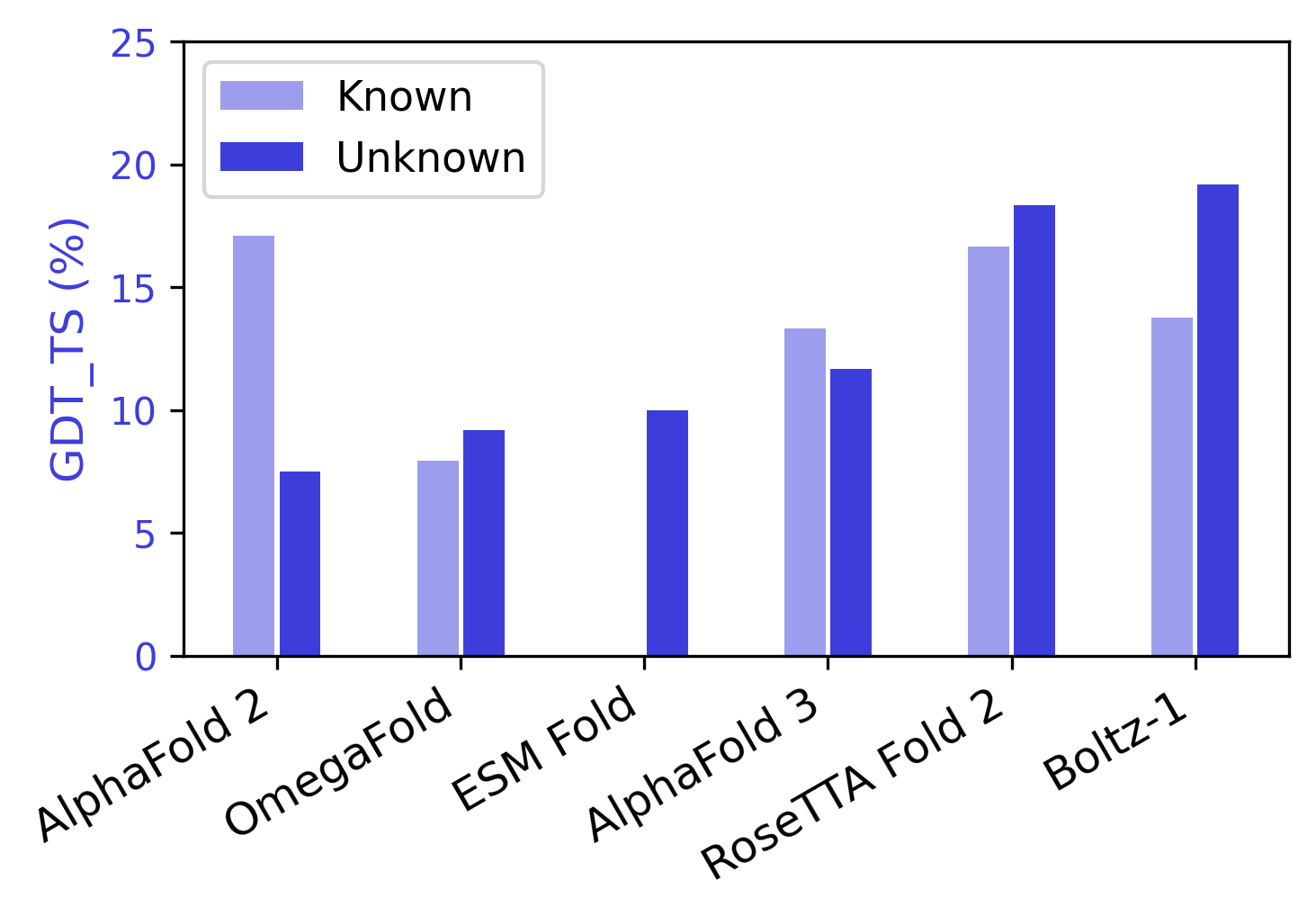}
        \caption{GDT-TS}
        \label{fig:sub1}
    \end{subfigure}
    \hfill
    \begin{subfigure}{0.49\textwidth}
        \centering
        \includegraphics[width=\textwidth]{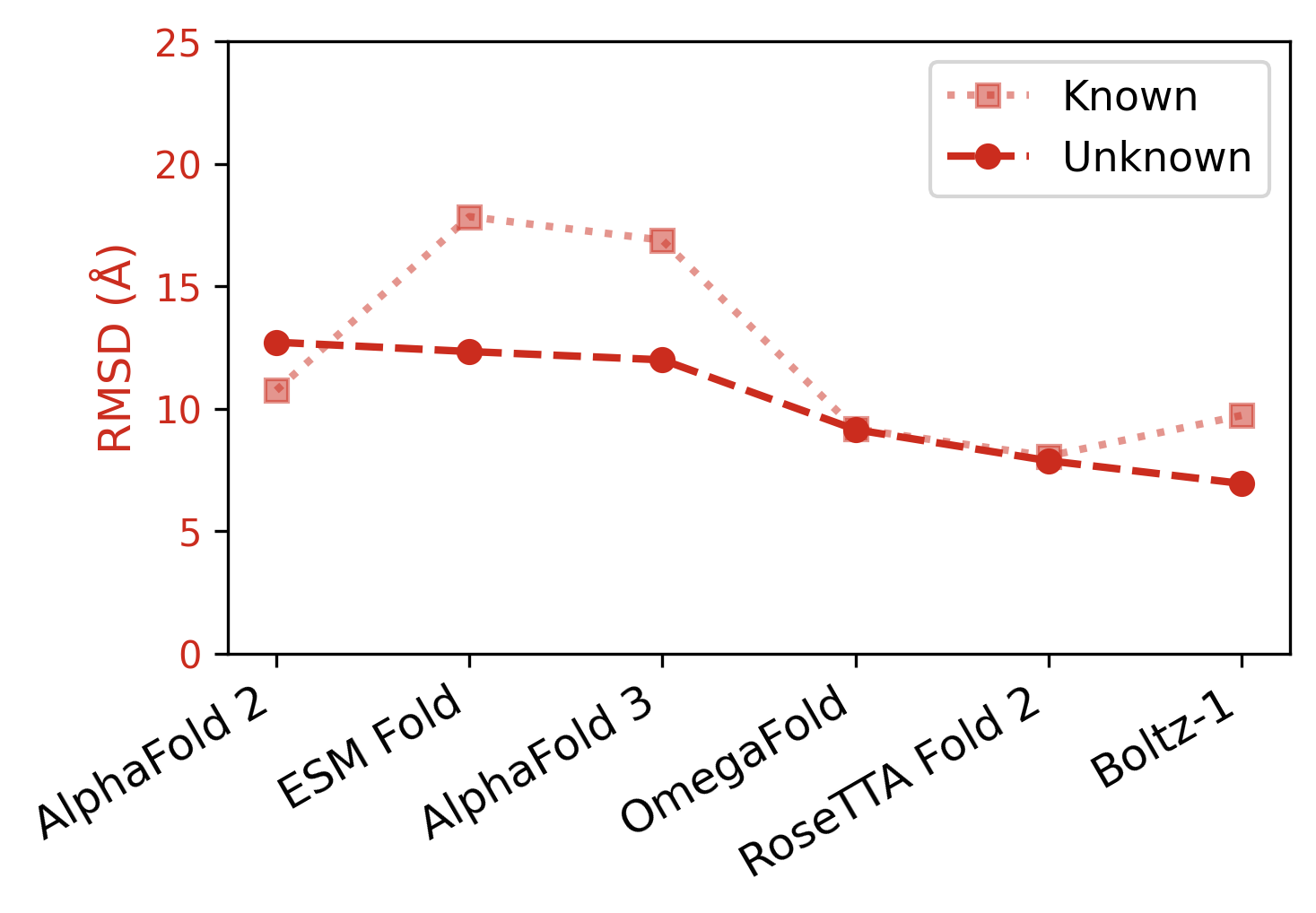}
        \caption{RMSD}
        \label{fig:sub2}
    \end{subfigure}
    \caption{Experimental results. We report metrics for proteins with an experimentally determined 3D structure (`known'), and out-of-domain sequences without a known structure (`unknown').}
    \label{fig:results}
\end{figure}

We run structure prediction models for the 10 sactipeptide sequences listed in table \ref{tab:sactipeptides} using 6 recent protein models: AlphaFold 2 \citep{jumper2021}, Boltz-1 \citep{wohlwend2024boltz1}, OmegaFold \citep{wu2022omegafold}, and RoseTTA Fold 2 \citep{Baek2023}, for which we use the ColabFold MMseqs2 MSA webserver by \cite{mirdita2022}; AlphaFold 3, using the DeepMind webserver \citep{alphafold3}; and ESM Fold, using the Meta ESM Atlas server \citep{lin2023}. We specify random seed 42 when possible for reproducibility purposes, and use the structure with highest reported confidence for models with ensembling. We compute RMSD and GDT-TS for all models on all known sactibonds, and report the results in table \ref{tab:results} and figure \ref{fig:results}.

\begin{table}[!t]
    \centering
    \begin{tabular}{llc|r r|r r}
         \textbf{Model} &  \textbf{Authors}  & \textbf{Ensembling} &  \multicolumn{2}{c|}{\textbf{GDT-TS}} & \multicolumn{2}{c}{\textbf{RMSD}} \\
          & & & Known & Unknown  & Known & Unknown \\
          \hline 
         AlphaFold 2 & DeepMind & Yes & 17.1 \% & 7.5 \% & 10.7 \AA & 12.7 \AA \\
         AlphaFold 3 & DeepMind &  Yes & 13.3 \% & 11.7 \% & 16.9 \AA & 12.0 \AA \\
         Boltz-1 & MIT & No & 13.7 \% & \textbf{19.2 \%} & 9.7 \AA & \textbf{6.9 \AA} \\
         ESMFold & Meta & No & 0.0 \% & 10.0 \% & 17.8 \AA & 12.3 \AA \\
         OmegaFold & Tencent AI & No &  7.9 \% & 9.2 \% & 9.2 \AA & 9.1 \AA \\
         RoseTTAFold 2 &  Baker Lab & Yes & 16.7 \% & 18.3 \% & 8.1 \AA & 7.9 \AA \\ \hline
         Average & -- & -- & 11.45 \% & 12.65 \% & 12.1 \AA & 10.1 \AA \\
    \end{tabular}
    \caption{Crosslinks structure prediction task results for proteins with experimentally determined structure (`known'), and for out-of-domain sequences without a known structure (`unknown').}
    \label{tab:results}
\end{table}

\section{Results}

\subsection{Performance on Crosslink Structure Prediction Task}

Our main finding is that performance on sactipeptide crosslinks structure prediction is limited for all tested models, with an average GDT-TS on known sactibonds around 11.5\% and a mean RMSD of 12.1~\AA\ (table~\ref{tab:results}). Interestingly, performance improves marginally for out-of-domain sequences with no resolved structure, with GDT-TS at 12.6\% and RMSD at 10.1~\AA. Boltz-1 and RoseTTAFold~2 yield the highest GDT-TS score for unknown sactipeptides, suggesting partial capture of local geometry. However, none of the models consistently approaches the ground truth 1.8~\AA\ sulfur-to-\(\alpha\)-carbon bond length, underscoring critical gaps in handling rare post-translational modifications.

\subsection{Analysis of Predicted Structures}

Visual inspection reveals that most models link sulfur atoms to one another through disulfide bonds, ignoring thioether bridge crosslinks. For example, the ruminococcin C1 structure predicted by AlphaFold 3, our highest-scoring conformation (50\% GDT-TS), positions all 4 sulfur atoms in disulfide bonds. This implies heavy reliance on standard cysteine-cysteine pairing, which is interesting as the structures reported in the PDB for these sactipeptides all bear sactibonds. Models also often collapses sulfur atoms in unnatural and sterically hindered conformations, highlighting a systematic bias against thioether bonds, which we hypothesize stems from their sparsity in structural data.



\begin{figure}[!hb]
    \centering
    \begin{subfigure}{0.49\textwidth}
        \centering
        \includegraphics[width=\textwidth]{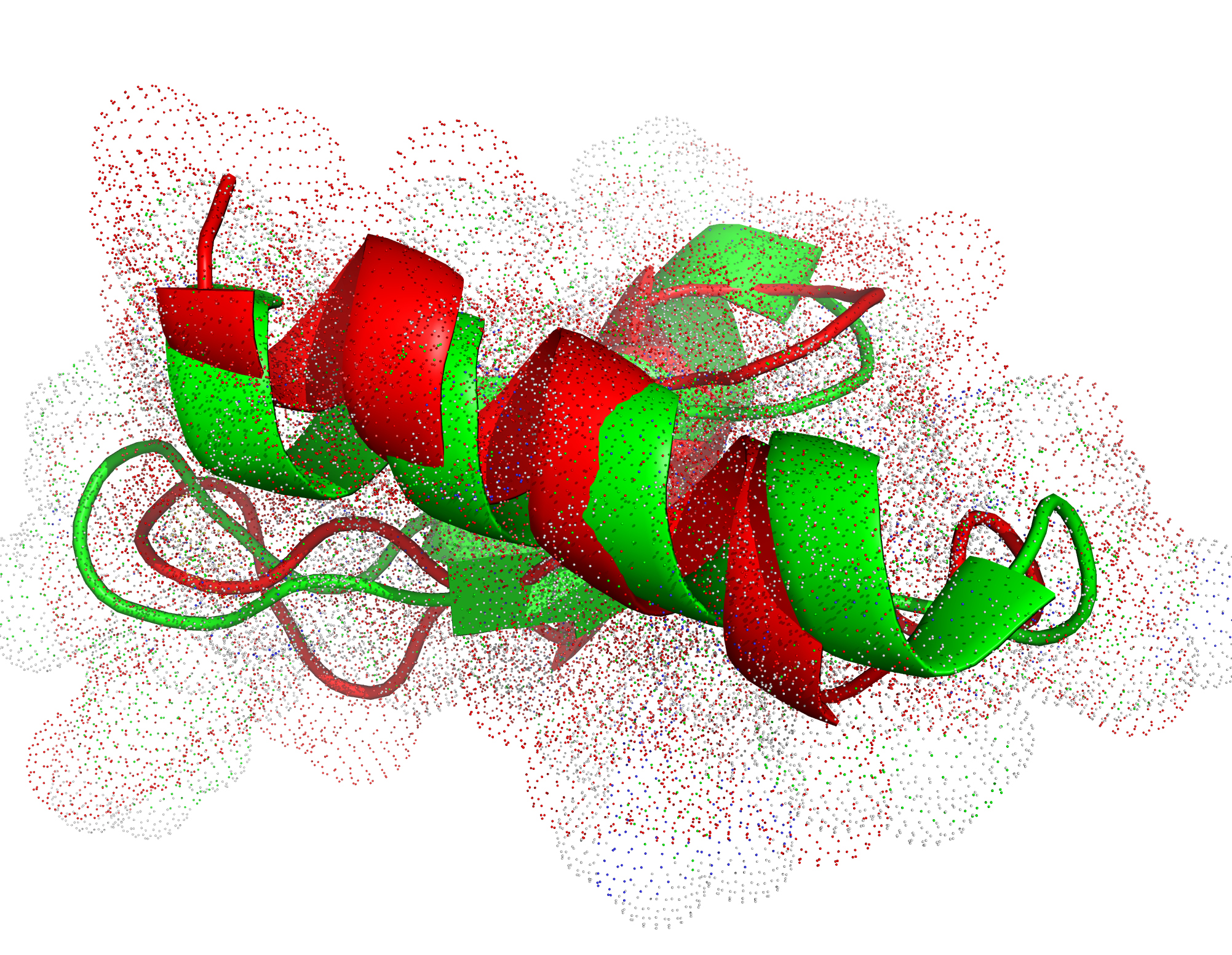}
    \end{subfigure}
    \hfill
    \begin{subfigure}{0.49\textwidth}
        \centering
        \includegraphics[width=\textwidth]{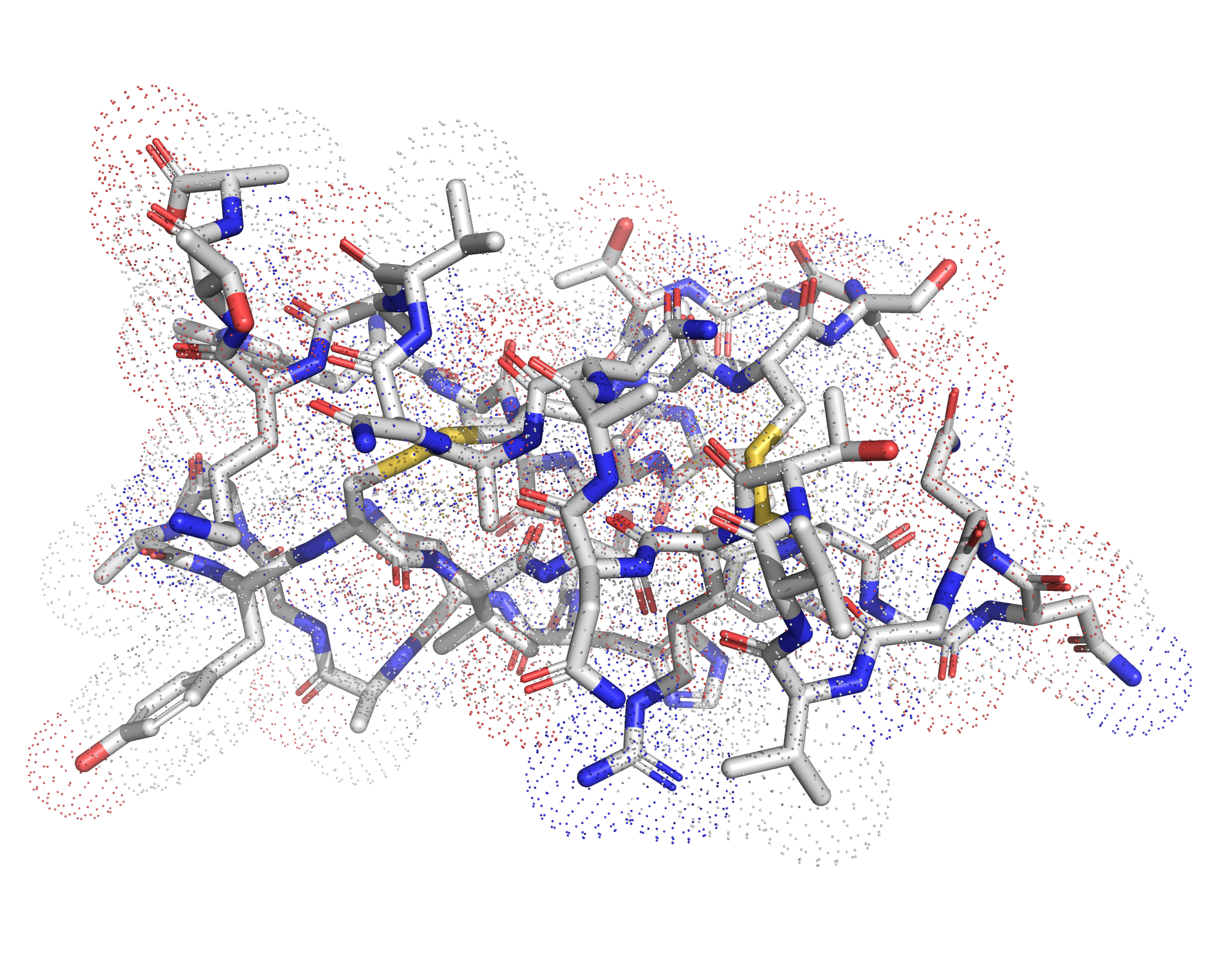}
    \end{subfigure}
    \caption{Ruminoccocin C1 structure predicted by AlphaFold 3: (a) super-imposed with the experimentally determined structure (left); (b) highlighting erroneously predicted disulfide bonds (right).}
    \label{fig:rumC}
\end{figure}

\subsection{Limitations and Future Work}

Our experiments are zero-shot and rely on pretrained models which weren't fine-tuned on sactipeptides. Substantial improvements may therefore emerge from incorporating explicit sulfur-to-\(\alpha\)-carbon constraints, or from training on curated examples of these rare PTMs. We also do not explore multi-conformation sampling, which might better capture sactibond geometry. Future research could investigate fusing physics-based or data-driven potentials with evolutionary amino-acid contact maps to improve robustness and generalization.



\section{Conclusion}
We presented an out-of-domain benchmark evaluating protein structure prediction models on sulfur-to-\(\alpha\)-carbon thioether bonds in sactipeptides. Our findings confirm that modern evolution-based predictors, including AlphaFold and RoseTTAFold, struggle with rare post-translational polycyclic structures. Although some models perform markedly better than other, including on unknown sactipeptides, none replicate accurate crosslink bonds lengths or structure geometry. These results underscore the need for physics-informed models to better predict rare post-translational modifications, and improve predictions of the 3D structure of biomolecules beyond evolutionary priors.





\newpage
\bibliography{iclr2025_conference}
\bibliographystyle{iclr2025_conference}
\end{document}